# Synthesis-related nanoscale defects in Mo-based Janus monolayers revealed by cross-correlated AFM and TERS imaging

Tianyi Zhang[1, #], Andrey Krayev[2, #], Tilo H. Yang[1], Nannan Mao[1], Lauren Hoang[3], Zhien Wang[1], Hongwei Liu[1], Yu-Ren Peng[1,4], Yunyue Zhu[1], Eleonora Isotta[5], Maria E. Kira[6], Ariete Righi[6], Marcos A. Pimenta[6], Yu-Lun Chueh[1,4], Eric Pop[3,7], Andrew J. Mannix[7], and Jing Kong[1]*

Tianyi Zhang, Tilo H. Yang, Nannan Mao, Zhien Wang, Hongwei Liu, Yu-Ren Peng, Yunyue Zhu, Jing Kong
Department of Electrical Engineering and Computer Science, Massachusetts Institute of Technology, Cambridge, Massachusetts 02139, United States
E-mail: jingkong@mit.edu

Andrey Krayev
HORIBA Scientific, Novato, California, 94949, United States

Lauren Hoang, Eric Pop
Department of Electrical Engineering, Stanford University, Stanford, California 94305, United States

Yu-Ren Peng, Yu-Lun Chueh
Department of Materials Science and Engineering, National Tsing Hua University, Hsinchu, 30013, Taiwan

Eleonora Isotta
Max Planck Institute for Sustainable Materials, Dusseldorf 40237, Germany

Maria E. Kira, Ariete Righi, Marcos A. Pimenta
Departamento de Física, Universidade Federal de Minas Gerais, Av. Antônio Carlos, 6627, Pampulha, Belo Horizonte, 31270-901, Minas Gerais, Brazil

Eric Pop, Andrew J. Mannix

Department of Materials Science & Engineering, Stanford University, Stanford, California 94305, United States

Keywords: 2D materials; Janus transition metal dichalcogenides; tip-enhanced Raman spectroscopy; atomic force microscopy; nanoscale defects

**Abstract**

Two-dimensional (2D) Janus transition metal dichalcogenides (TMDs) are promising candidates for various applications in non-linear optics, energy harvesting, and catalysis. These materials are usually synthesized via chemical conversion of pristine TMDs. Nanometer-scale characterization of the obtained Janus materials' morphology and local composition is crucial for both the synthesis optimization and the future device applications. In this work, we present a cross-correlated atomic force microscopy (AFM) and tip-enhanced Raman spectroscopy (TERS) study of Janus $Mo^S_{Se}$ and Janus $Mo^{Se}_S$ monolayers synthesized by the hydrogen plasma-assisted chemical conversion of $MoSe_2$ and $MoS_2$, respectively. We demonstrate how the choice of the growth substrate and the starting TMD affects the morphology of the resulting Janus material. Furthermore, by employing TERS imaging, we demonstrate the presence of nanoscale islands (~20 nm across) of $MoSe_2$- $Mo^S_{Se}$ ($MoS_2$-$Mo^{Se}_S$) vertical heterostructures originating from the bilayer nanoislands in the precursor monolayer crystals. The understanding of the origins of nanoscale defects in Janus TMDs revealed in our study can help with further optimization of the Janus conversion process towards uniform and wrinkle-/crack-free Janus materials. Moreover, our work shows that cross-correlated AFM and TERS imaging is a powerful and accessible method for studying nanoscale composition and defects in Janus TMD monolayers.

**Introduction**

Two-dimensional (2D) Janus transition metal dichalcogenides (TMDs) have an asymmetric structure of $X_1$-M-$X_2$, where a layer of transition metal atoms (denoted as M) is sandwiched between two different chalcogen atom layers (denoted as $X_1$ and $X_2$).[1] Several intriguing properties arise from this unique structure, such as intrinsic vertical dipole and strain anisotropy.[2] These emergent properties, coupled with the inherent atomic thickness and semiconducting characteristics, have made Janus TMDs a class of interesting two-dimensional (2D) materials for both the studies of their fundamental properties and prospective applications in non-linear optics, photovoltaics, piezotronics, catalysis, etc.[3]

Janus TMDs do not exist in the bulk crystal form, neither man-made nor naturally occurring. Thus, establishing a well-controlled bottom-up synthetic route is indispensable for the manufacturing of these monolayers. So far, the synthesis of Janus TMDs has been realized by selectively substituting one of the chalcogen layers in the TMD precursor with a different type of chalcogen. For example, using monolayer $MoSe_2$ grown on gold foils as the starting material, the bottom selenium layer at the $MoSe_2$-gold interface was selectively replaced with sulfur at high temperature (~700º C), forming Janus $Mo_S^{Se}$ (here Se/S are top/bottom chalcogen layers with respect to the substrate).[4] In a different study, a controlled low-energy selenium implantation method was used to replace the top sulfur layer in monolayer $WS_2$ leading to the formation of Janus $W_S^{Se}$ [5]. Additionally, a novel room-temperature remote hydrogen ($H_2$) plasma-assisted method was developed for the replacement of the top chalcogen layers in TMD crystals grown on $Si/SiO_2$, thus realizing a low-temperature approach for the Janus TMD synthesis.[6] Existing reports have demonstrated that the remote $H_2$ plasma-assisted Janus conversion is capable of both converting $MS_2$ into Janus $M_S^{Se}$, and converting $MSe_2$ into Janus $M_{Se}^{S}$, seemingly obtaining the same Janus TMD material, but with inversed order of the corresponding chalcogen layers and, consequently, opposite electric dipole orientation.[6-7] Nevertheless, the composition, quality, and/or morphology of these two seemingly identical reaction products have not yet been investigated in detail. To achieve this goal, reliable characterization methods with nanometer-scale spatial resolution are required. While conventional optical microscopy techniques, such as far-field Raman and photoluminescence (PL) spectroscopy, may be rapid approaches to investigate strain, doping, and defects in 2D materials,[8] these techniques cannot capture fine nanoscale structural features due to the diffraction limit constraint of their spatial resolution. On the other hand, atomic-resolution characterization approaches, such as scanning transmission electron microscopy (STEM) and scanning tunneling microscopy (STM), can be employed to visualize atomic-scale defects and local strain,[9] but are very demanding towards the sample preparation and are not very practical for capturing larger-area information (50-1000 nm) on the morphology and uniformity of the investigated materials.[10] Moreover, these methods inevitably involve the transfer of Janus TMDs to specific substrates, which still remains rather challenging due to the intrinsic strain within Janus TMD crystals which leads to a strong tendency of forming nanoscrolls, in turn preventing the characterization of the properties of flat Janus crystals.[2a]

In our current study, we used cross-correlated atomic force microscopy (AFM) and tip-enhanced Raman spectroscopy (TERS) imaging to study the morphology and chemical composition of Janus $Mo_{Se}^{S}$ and Janus $Mo_{S}^{Se}$ monolayers synthesized by $H_2$ plasma-assisted conversion of chemical vapor deposition (CVD)-grown $MoSe_2$ and $MoS_2$, respectively. We found that the two conversion pathways induce distinct strain which may lead to the formation of either nanoscale cracks or wrinkles in synthesized Janus TMDs. In addition, we demonstrated that TERS characterization is very effective in revealing nanoscale features, such as nano-islands of $MoSe_2$-Janus $Mo_{Se}^{S}$ vertical heterostructures of only 15 ~20 nm across, as well as differentiating the cracks from inverted nano-wrinkles. Our work provides a solid experimental approach towards the characterization of the nanoscale composition and defects in Janus TMDs, and reveals physical origins of these defects, thus outlining possible approaches towards the synthesis of perfect cracks- and wrinkles-free Janus TMD monolayers.

**Results and discussion**

The synthesis of MoSSe-type Janus monolayers can start either from $MoSe_2$ or $MoS_2$ precursors with consequent sulfurization/selenization of the topmost atomic layer. Using the $H_2$ plasma-assisted atomic-layer substitution (ALS) method (see Methods), we successfully implemented both reaction pathways, resulting in the $Mo_{Se}^{S}$ and $Mo_{S}^{Se}$ monolayers correspondingly (see optical images in **Figures 1a-b**). At first glance, Janus monolayers obtained via these two routes should be identical, except for their opposite atomic layer order in Z direction. However, it turns out not to be the case, as evidenced by the far-field Raman spectroscopy characterization using 532 nm excitation (**Figure 1c**). Although both $Mo_{Se}^{S}$ and $Mo_{S}^{Se}$ display characteristic out-of-plane $A_1^1$ and in-plane $E^2$ modes associated with Janus MoSSe[11], the frequency of in-plane $E^2$ mode displays a difference of > 4 $cm^{-1}$ between these two types of Janus monolayers. As the in-plane Raman mode in 2D TMDs is sensitive to in-plane strain[8a], the observed frequency shift of $E^2$ mode suggests a difference in residual strain in Janus $Mo_{Se}^{S}$ and $Mo_{S}^{Se}$ crystals. The strain in Janus monolayers is induced during both the TMD growth and the ALS reaction stages, but the degree and even the type of final strain in $Mo_{Se}^{S}$ and $Mo_{S}^{Se}$ should differ, because at the ALS reaction stage they experience different lattice expansion/contraction processes. More specifically, as illustrated in **Figures 1d-e**, based on our theoretical estimates from Raman spectra (see **Figure S1**), the monolayer $MoS_2$ and $MoSe_2$ grown on $SiO_2$/Si should acquire a tensile strain of ~0.6% due to the strong interaction (pinning) between the TMD crystal and the growth substrate, and the significant mismatch between the thermal expansion coefficients (TECs) of the TMD and the substrate.[12] The ALS conversion process further introduces strain due to the lattice constant change, and here we use literature reported lattice constant values to obtain the strain estimations.[2b, 13] In the case of converting $MoSe_2$ to $Mo_{Se}^{S}$, pre-existing tensile strain is complemented by additional tensile strain caused by sulfurization of the top layer (the lattice constant is decreased from $\alpha_{MoSe_2} = 3.288$ Å to $\alpha_{MoSSe} = 3.23$ Å), leading to the final combined value of ~2.4% (**Figure 1d**). In contrast, upon conversion of $MoS_2$ to $Mo_{S}^{Se}$, due to the replacement of the top sulfur atomic layer with selenium, a ~2.2% compressive strain should appear (the lattice constant is increased from $\alpha_{MoS_2} = 3.16$ Å to $\alpha_{MoSSe} = 3.23$ Å), but this compressive strain is partially compensated by the original tensile strain, leading to residual compressive strain value in $Mo_{S}^{Se}$ crystals of ~1.6% (**Figure 1e**). We would like to stress again that the seemingly counterintuitive sign of strain in response to the lattice constant increase/decrease at the conversion stage is caused by the strong pinning of the TMD and later- Janus TMD crystals to the substrate.

Having understood the in-plane strain difference in Janus monolayer $Mo_{Se}^{S}$ and $Mo_{S}^{Se}$ converted from $MoSe_2$ and $MoS_2$, respectively, we can now focus on the comparison of the corresponding crystal morphologies. Janus TMD crystals were assessed by cross-correlated AFM and TERS imaging (see **Methods** for details on the synthesis of Janus TMDs) on both the growth substrates and after a gold- or silver-assisted transfer. The latter transfer from the growth substrate ($SiO_2$/Si or fused silica) was performed in order to maximize TERS response by enabling the so-called gap mode TERS conditions. Here we followed the previously developed metal-assisted dry-transfer procedure (**Figure S2a**).[14] This transfer process results in Janus TMDs embedded in the gold or silver substrate, which, as mentioned already, enables the optimal gap-mode conditions for TERS characterization, as well as the efficient use of Kelvin probe microscopy and photocurrent imaging techniques. It is worth noting that this transfer process involves the mechanical delamination and flipping of Janus TMDs, thus exposing the side of the crystal that was originally facing the growth substrate. The schematics of Janus Mo$_{Se}^{S}$ monolayer and the Mo$_{Se}^{S}$-MoSe$_2$ heterobilayer before and after transfer is illustrated in **Figure S2**.

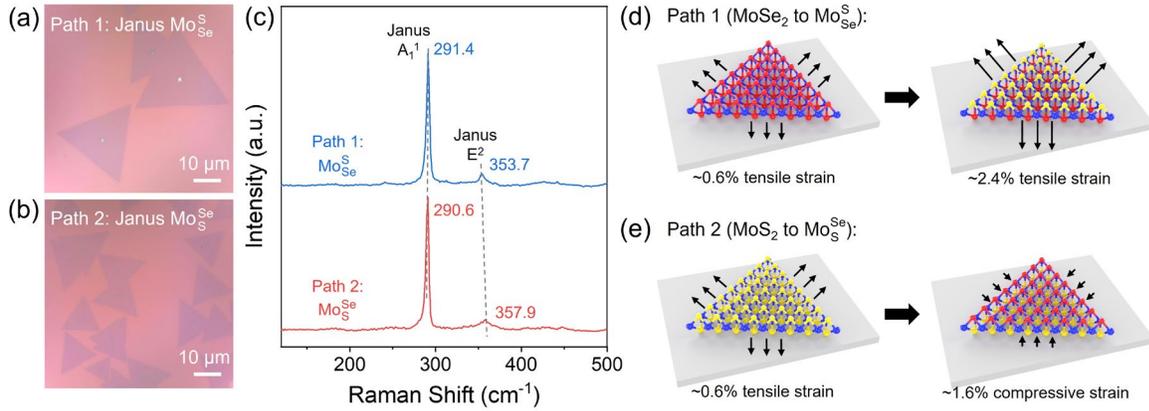

**Figure 1**. **Two reaction pathways of obtaining MoSSe-type Janus monolayers via either the sulfurization of MoSe$_2$ or the selenization of MoS$_2$.** (a-b) Optical microscopy images of as-synthesized Janus Mo$_{Se}^{S}$ and Mo$_{S}^{Se}$ on the SiO$_2$/Si substrate. (c) Far-field Raman spectroscopy characterization of as-grown Janus Mo$_{Se}^{S}$ and Mo$_{S}^{Se}$. (d-e) Schematics illustrating that the two reaction pathways introduce different types and amount of strain into the resulting Janus monolayers.

We first start with the results obtained on Mo$_{Se}^{S}$ converted from MoSe$_2$. The as-grown MoSe$_2$ flakes were primarily triangular monolayer crystals with lateral sizes of 20 μm and more (**Figure 2a**). From scanning transmission electron microscopy high angle annular dark field (STEM-HAADF) characterization, we observed the presence of nanoscale bilayer MoSe$_2$ islands with lateral sizes of ~20 nm (**Figure 2b**), and the majority of these islands were 3R-stacked (rhombohedral)[15], as confirmed by the atomic-resolution STEM-HAADF image in **Figure 2c**. Upon conversion to Janus Mo$_{Se}^{S}$, the resulting crystals fragmented into smaller, approximately 500-1000 nm-across, physically separated domains due to excessive tensile strain introduced into the lattice, as evidenced by the AFM topography characterization (**Figure S3**) as well as surface potential and TERS characterization that will be discussed in the following part of the manuscript.

We performed Kelvin probe microscopy and TERS imaging (785 nm excitation) of Janus Mo$_{Se}^{S}$ transferred to gold. Contact potential difference (CPD) images in **Figure 2d** clearly show the presence of small domains, which are similar in size and shape to the ones observed in the AFM topography of as-synthesized Janus Mo$_{Se}^{S}$ crystals (**Figure S3**). A TERS map was further collected over the junction of two adjacent domains (**Figures 2e-f**) to reveal detailed chemical composition information on both the domains and the junction lines. The colors in the combined TERS map (**Figure 2f**) represent the intensities of $A_1$'(MoSe$_2$), $A_1^2$(Mo$_{Se}^{S}$), and $A_1^1$(Mo$_{Se}^{S}$) vibrational modes, respectively. From the representative averaged TERS spectra (blue-colored spectrum in **Figure 2g**) collected over the main body of the domain (marked with the blue rectangle in **Figure 2f**), we see a clear signature of Janus monolayer Mo$_{Se}^{S}$ with characteristic $A_1^1$ mode at ~289 cm$^{-1}$ and $A_1^2$ at ~429 cm$^{-1}$. The Raman modes of the starting MoSe$_2$ are absent, indicating a full conversion to Janus Mo$_{Se}^{S}$. Interestingly, the Mo$_{Se}^{S}$ $A_1^2$ band is more intense than $A_1^1$, and is also noticeably split. We observed no characteristic Raman modes in the cracks between the adjacent domains (green-colored spectra in **Figure 2g**), which confirms that these domains are physically separated with ~30 nm wide gaps. Finally, we found small islands with sizes of 20-30 nm (marked with the red square in **Figure 2f**) which showed simultaneously strong intensity of $A_1$' mode of MoSe$_2$ at ~242 cm$^{-1}$ and the typical Raman modes of Janus Mo$_{Se}^{S}$, which is a spectroscopic signature of the Janus Mo$_{Se}^{S}$-MoSe$_2$ vertical heterostructures (schematically shown in **Figure S2c**). This correlates very well with the STEM observation of the bilayer MoSe$_2$ nano-islands in the precursor material, as the self-limiting ALS method converts only the topmost Se layer into S, resulting in formation of Mo$_{Se}^{S}$ only in the top layer of the bilayer MoSe$_2$. It should be noted again that conventional far-field Raman microscopy of the as-grown Janus TMDs is not able to resolve these nanoscale heterostructures or cracks due to insufficient spatial resolution (**Figure S4**). This further highlights the immense potential of using TERS imaging to capture detailed morphological and chemical information of Janus TMDs with relevant, nm-scale spatial resolution.

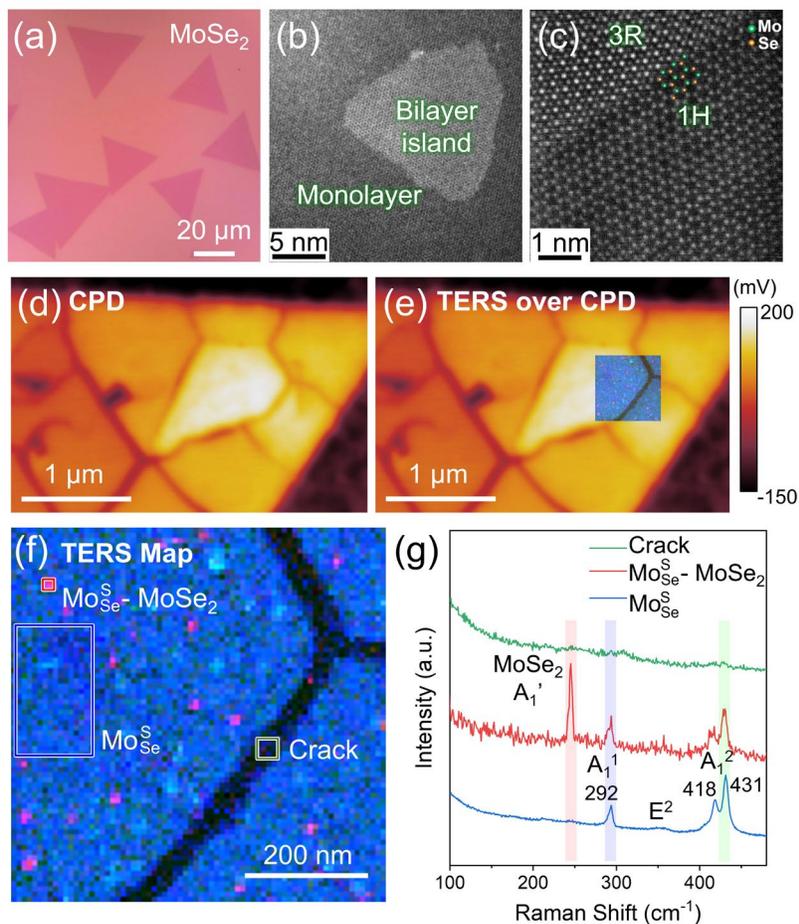

**Figure 2. AFM and TERS characterization of Janus $Mo^S_{Se}$ converted from $MoSe_2$.** (a) Optical microscopy image of CVD-grown $MoSe_2$ on $SiO_2$/Si which is used as the precursor for the ALS conversion. (b) Low-magnification and (c) atomic-resolution STEM-HAADF images of monolayer and bilayer regions of CVD-grown $MoSe_2$. (d) Contact Potential Difference (CPD) image of the transferred Janus $Mo^S_{Se}$ flake on the gold substrate. (e) The TERS map overlaid over the CPD image of Janus $Mo^S_{Se}$. (f) The TERS map showing the intensity distribution of correspondingly highlighted Raman modes in panel (g), namely - $MoSe_2$ $A_1'$ (red), $Mo^S_{Se}$ $A_1^2$ (green), and $Mo^S_{Se}$ $A_1^1$ (blue). (g) The TERS spectra averaged over the correspondingly colored boxes in panel (f), which represent Janus monolayer $Mo^S_{Se}$ (blue), $Mo^S_{Se}$-$MoSe_2$ vertical heterostructures (red), and crack (green) regions.

To further confirm the multi-layer nature of the nano-islands in Janus $Mo^S_{Se}$ crystals, we collected another TERS map on an adjacent domain. In this domain, we recorded the photocurrent generated across the crystal (along Z direction, between the TERS probe and the substrate) together with the TERS spectra in each pixel of the map (**Figure S5**). Comparing the TERS and photocurrent maps in **Figures S5e and f**, we can see that the photocurrent greatly increases over the nanoislands that were showing the spectral signatures of both the Janus $Mo^S_{Se}$ and starting

MoSe$_2$. Enhanced photocurrent should be expected in the vertical heterostructures, because for 2D TMDs with a few-layer thicknesses, the photocurrent value is limited by the absorption in the material which grows with the number of layers, at least within the low layer number limit. The exact scaling of the photocurrent with the layer number of the TMD crystals may vary depending on the excitation wavelength, showing almost an exponential growth under illumination slightly below the A exciton energy in the case of few-layer WS$_2$ (**Figure S6**).

Now we switch to the results obtained on Janus Mo$_S^{Se}$ crystals converted from MoS$_2$. In order to understand the morphological features of Janus Mo$_S^{Se}$, we first collected the AFM topography image of a specific crystal on the growth substrate (**Figure 3a**) and then located the same crystal after the silver-assisted transfer to perform another AFM characterization (**Figure 3b**). We confirmed that the two AFM images were obtained on the exact same area by superimposing the before- (the image is flipped to properly represent the mirror inversion after the transfer) and after-the-transfer topography (**Figure S7**), that produced a perfect match. From the topography image in **Figure 3a**, wrinkles with typical heights of ~5 nm are observed on as-grown Janus Mo$_S^{Se}$. We assign their appearance to the compressive strain introduced into the lattice during the Janus conversion of MoS$_2$, as we discussed earlier (**Figure 1** and related discussion). Our observations confirm that the different types and the amount of strain induced during the ALS conversion stage can lead to fundamentally different morphologies in resulting Janus TMDs.

Though the wrinkles got inverted upon the silver-assisted transfer process and looked like the cracks in Janus Mo$_{Se}^{S}$ crystals (**Figure 3b**), TERS imaging clearly differentiated these inverted wrinkles from the cracks. **Figure 3c** presents a magnified AFM topography image of the region highlighted by the white square in **Figure 3b**. A TERS map was collected from the white square-marked region in **Figure 3c**, and is shown in **Figure 3d**. Similar to the TERS image of Janus Mo$_{Se}^{S}$ discussed earlier, the colors in TERS map in **Figure 3d** represent the intensities of A$_1$'(MoS$_2$)-red, A$_1^2$(Mo$_S^{Se}$)-green, and A$_1^1$(Mo$_S^{Se}$)-blue modes, correspondingly. From both the TERS map (regions marked with the green rectangle) and the individual TERS spectra (**Figure 3e**), we can clearly see that the TERS signal of Janus Mo$_S^{Se}$ enhances over the inverted wrinkles instead of vanishing, confirming the presence of wrinkles rather than cracks. We also observed islands of Janus Mo$_S^{Se}$-MoS$_2$ vertical heterostructures (~100 nm across, as marked with the red rectangle in **Figure 3d**), which is evidenced by the strong A$_1$' mode of MoS$_2$ at ~405 cm$^{-1}$ and the typical Raman modes of

Janus Mo$_S^{Se}$. Notably, while the microscopic islands of vertical heterostructures of Janus Mo$_S^{Se}$-MoS$_2$ were captured both by the Kelvin probe microscopy and TERS imaging, the nanoscale islands were only detected by TERS imaging (**Figure S8**), which again highlights the power of this technique for the nanoscale characterization of Janus TMDs and other 2D semiconductors. The reason for such seemingly counterintuitive discrepancy between the spatial resolution of the Kelvin probe and TERS maps is related to the fact that in Kelvin measurements the spatial resolution is limited by the tip radius (50-100 nm for TERS probes) and the tip-sample separation in the Kelvin pass (20-30 nm), while in TERS measurements performed de facto in contact mode it's the thickness of the TMD layer (~1-2 nm) that separates the tip and the substrate which is the scaling factor for the spatial resolution.

Additionally, we noticed significant differences in TERS spectra of Janus monolayer Mo$_S^{Se}$ with Mo$_{Se}^S$ crystals. Specifically, in Mo$_S^{Se}$, the $A_1^1$ mode is more intense than $A_1^2$ (blue-colored spectra in **Figure 3e**), which is the opposite of what we observed in Mo$_{Se}^S$ (blue-colored spectra in **Figure 2g**). Moreover, $A_1^1$ is obviously split in Janus Mo$_S^{Se}$, when compared to the case of the $A_1^2$ peak splitting in the TERS spectrum of Janus Mo$_{Se}^S$. The possible mechanism that leads to these differences will be discussed later in the manuscript.

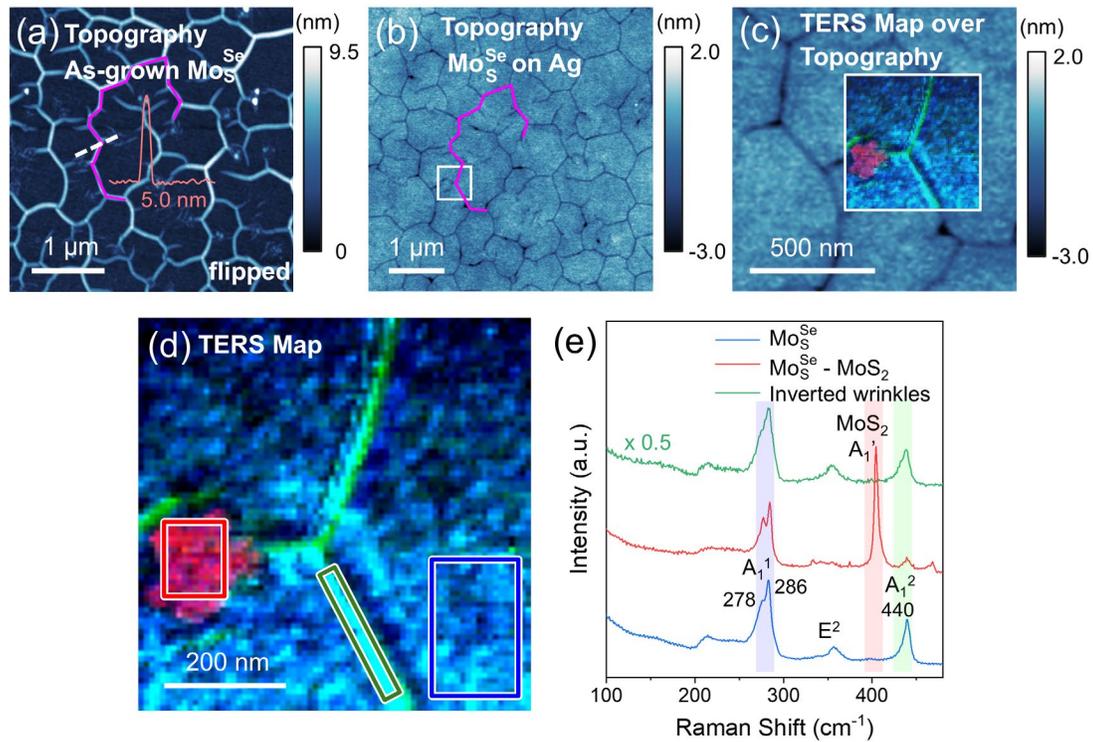

**Figure 3. AFM and TERS characterization of Janus Mo$_S^{Se}$ converted from MoS$_2$.** (a) AFM topography image (image flipped to compare the mirror inversion after the transfer) of as-grown Janus Mo$_S^{Se}$ on SiO$_2$/Si. Inset: an AFM line profile of a typical wrinkle in Janus Mo$_S^{Se}$, showing a height of 5.0 nm. (b) AFM topography image of the same region of the Janus Mo$_S^{Se}$ crystal after being transferred onto silver. (c) Magnified AFM topography image of the region highlighted by the white square in (b). The TERS map is also acquired from this region, and is superimposed on the topography image. (d) The magnified TERS map showing the intensity distribution of correspondingly highlighted Raman modes in panel (e), namely - MoS$_2$ A$_1$' (red), Mo$_S^{Se}$ A$_1^2$ (green), and Mo$_S^{Se}$ A$_1^1$ (blue). (e) The TERS spectra averaged over the correspondingly colored boxes in panel (d), which represent Janus monolayer Mo$_S^{Se}$ (blue), Mo$_S^{Se}$-MoS$_2$ vertical heterostructures (red), and inverted wrinkle (green) regions.

While the wrinkle formation within the Janus TMD crystals may open up their potential applications in sensing and catalysis, the ability to yield Janus TMDs with a flat and uniform surface morphology is desirable for prospective electronic and photonic applications. To attain this goal, our rationale is to reduce the amount of compressive strain in the synthesized Janus Mo$_S^{Se}$ by substrate engineering to mitigate wrinkle formation. In **Figures 1d and e**, the amount of strain induced during the TMD synthesis process is tunable by selecting substrates with different TECs[16], while the strain induced during the MoS$_2$ → Janus Mo$_S^{Se}$ conversion process is intrinsically determined by their lattice constant differences. Thus, we focused on using alternative substrates other than SiO$_2$/Si to introduce a larger amount of tensile strain in the precursor MoS$_2$, which can better compensate for the compressive strain during the Janus conversion process, decreasing the amount of remaining strain in Janus Mo$_S^{Se}$. To enhance the tensile strain in CVD-grown MoS$_2$, a low TEC substrate is desirable.[16] Here we selected fused silica with a very low TEC value of 0.55 × 10$^{-6}$ K$^{-1}$ as a suitable substrate. In comparison, the TEC of SiO$_2$/Si is expected to be governed by the TEC of Si (i.e., 2.7 × 10$^{-6}$ K$^{-1}$ at room temperature)[17], due to the much smaller thickness of the SiO$_2$ layer. Thus, a larger TEC mismatch between fused silica with MoS$_2$ (measured to be 7.6 ± 0.9 × 10$^{-6}$ K$^{-1}$)[18] can lead to increased tensile strain in CVD-grown MoS$_2$. In CVD-synthesized monolayer MoS$_2$ on fused silica, we observed red shifts in the MoS$_2$ E' Raman mode and PL emission compared to MoS$_2$ synthesized on SiO$_2$/Si (**Figure S9**). This further confirms the presence of increased tensile strain in MoS$_2$, as tensile strain leads to phonon softening and bandgap reduction, which induces red shifts in both Raman E' mode and PL emission energy.[12c, 19]

Similar to the case of Si/SiO$_2$ substrates, we synthesized Janus Mo$_S^{Se}$ on fused silica and characterized the same flake both as-grown and after the silver-assisted transfer (**Figures 4 and S9**). Note that we chose silver as the transfer medium and TERS substrate in order to extend the spectral range of TERS measurements to 473 nm, while still preserving TERS enhancement in NIR optical range.[20] From the topography image of as-synthesized Mo$_S^{Se}$ sample grown on quartz in **Figure S10**, the major part of the crystal has very smooth topography, indicating that the use of fused silica substrate mitigates the wrinkle formation. A small dendrite-like region is also present in the crystal, which shows a surface potential difference in the CPD map (**Figure 4b**). TERS measurements confirmed that this region is the vertical heterostructure of Janus Mo$_S^{Se}$-MoS$_2$ (**Figures 4d and e**). In terms of the TERS spectra from both the monolayer Janus Mo$_S^{Se}$ and Janus Mo$_S^{Se}$-MoS$_2$ vertical heterostructures (**Figure 4e**), they appeared to be very similar to the spectra from the corresponding crystals grown on SiO$_2$/Si (**Figure 3e**), indicating a comparable quality of Janus Mo$_S^{Se}$ and vertical heterostructures grown on fused silica and Si/SiO$_2$.

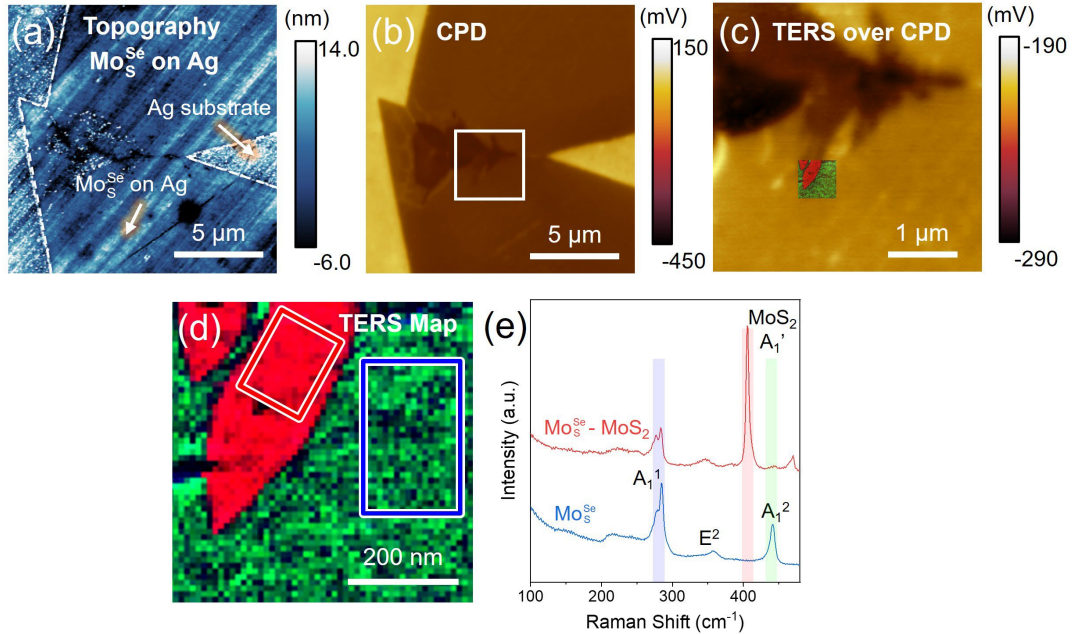

**Figure 4. AFM and TERS characterization of wrinkle-free Janus $Mo^{Se}_S$ synthesized on the fused silica substrate.** (a) AFM topography image of Janus $Mo^{Se}_S$ transferred from fused silica to silver. The edges of $Mo^{Se}_S$ are outlined by white dashed lines to differentiate $Mo^{Se}_S$ and bare silver substrate regions. (b) CPD image of the identical Janus $Mo^{Se}_S$ flake. (c) Magnified CPD image of the region highlighted by the white square in (b). The TERS map is also acquired from this region, and is superimposed on the CPD image. (d) The magnified TERS map showing the intensity distribution of correspondingly highlighted Raman modes in panel (e), namely - $MoS_2$ $A_1'$ (red), $Mo^{Se}_S$ $A_1^2$ (green), and $Mo^{Se}_S$ $A_1^1$ (blue). (e) The TERS spectra averaged over the correspondingly colored boxes in panel (d), which represent Janus monolayer $Mo^{Se}_S$ (blue) and $Mo^{Se}_S$-$MoS_2$ vertical heterostructures (red).

Finally, we attempted to gain a better understanding on the nature of spectral splitting of $A_1^1$ and $A_1^2$ modes in Janus $Mo^S_{Se}$ and Janus $Mo^{Se}_S$ monolayers by excitation-dependent TERS characterization. We collected TERS maps of the same area in the Janus $Mo^{Se}_S$ sample on silver with the same TERS probe at 785 nm and 473 nm excitation. The TERS maps are presented in **Figures 5a and b**, and the individual spectra are extracted from the regions marked with blue and red rectangles, which correspond to monolayer Janus $Mo^{Se}_S$ and Janus $Mo^{Se}_S$-$MoS_2$ vertical heterostructures, respectively (**Figures 5c and d**). From the TERS spectra corresponding to monolayer Janus $Mo^{Se}_S$, we notice a significant increase of the relative intensity of $A_1^2$ mode compared to the $A_1^1$ in the spectrum collected with 473 nm excitation (**Figure 5d**). The difference in TERS spectra obtained using 785 nm and 473 nm excitations is likely due to the excitation laser wavelength dependence of Raman intensities, which should be valid for both the far-field Raman (**Figure S11**) and TERS.[11] In addition, the obvious split of the $A_1^1$ mode observed in TERS spectra collected with 785 nm excitation is not visible in TERS spectra of the same area collected with 473 nm, though it may be also a consequence of insufficient spectral resolution in the latter case. We should also note that in the case of 473 nm excitation, we observe both the out-of-plane $A_1'$ and the in-plane $E'$ modes of MoS$_2$ in Janus $Mo_S^{Se}$-MoS$_2$ vertical heterostructures marked with red rectangles in TERS maps in **Figure 5b**, while the spectrum collected with 785 nm features only $A_1'$ mode. The appearance/absence of Raman bands in 2D TMDs with varied excitation reflects various electronic resonances and intricate electron-phonon interactions in these materials, and thus we can speculate that the $A_1^1$ and $A_1^2$ splitting observed in our experiments may also be a consequence of resonant conditions that can lead to the rise of some bands at certain excitation which are not visible at different excitation wavelengths. We must state though that a firm, well-experimentally supported answer to the question of the nature of the $A_1^1$ and $A_1^2$ mode splitting as well as the obvious inversion of the intensity of these bands in $Mo_S^{Se}$ and $Mo_{Se}^S$ crystals require a focused dedicated study, which is beyond the scope of our current manuscript.

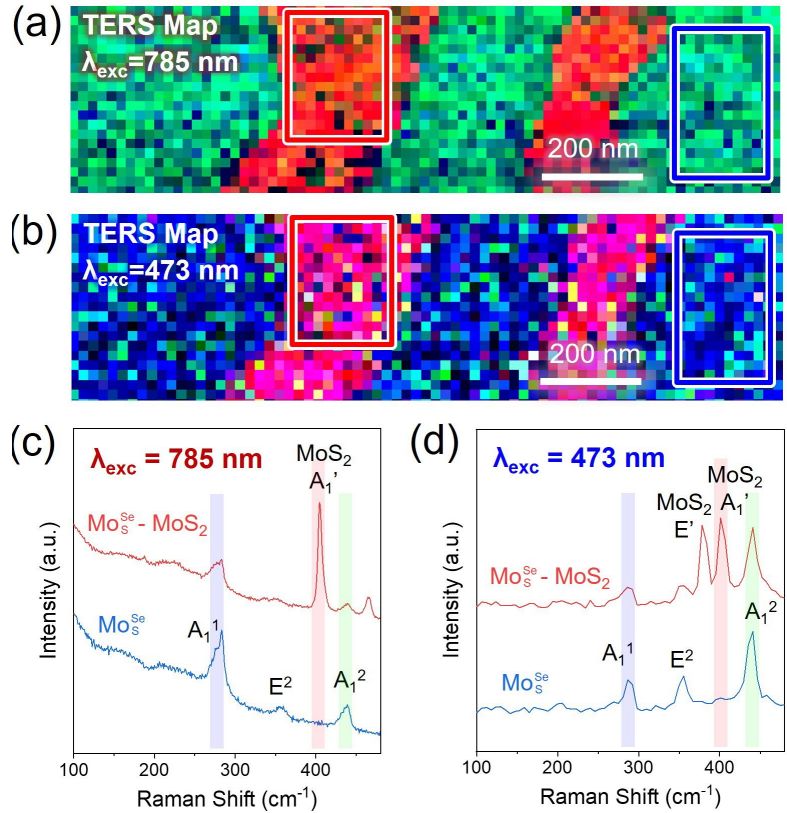

**Figure 5. Excitation-dependent TERS characterization of Janus Mo$_S^{Se}$ on silver.** (a-b) TERS maps of an identical region of Janus Mo$_S^{Se}$ using 785 nm and 473 nm excitation, respectively. The color of TERS maps indicates the intensity distribution of correspondingly highlighted Raman modes in panels (c-d), namely - MoS$_2$ A$_1$' (red), Mo$_S^{Se}$ A$_1^2$ (green), and Mo$_S^{Se}$ A$_1^1$ (blue). (c-d) The TERS spectra averaged over the correspondingly colored boxes in panels (a-b), which represent Janus monolayer Mo$_S^{Se}$ (blue) and Mo$_S^{Se}$-MoS$_2$ vertical heterostructures (red).

**Conclusion**

In this work, using cross-correlated AFM and TERS imaging, we investigated the typical synthesis-related nanoscale defects in Janus Mo$_{Se}^S$ and Mo$_S^{Se}$, monolayer crystals. We identified two principal types of defects - one originating from the presence of nano- to micro-scale multilayer islands in the predominantly monolayer TMD precursor crystals. Upon conversion, these islands become vertical heterostructures of the Janus material and the corresponding starting TMD underneath. The second type of defect is related to the type and the amount of residual strain incurred in crystals in the course of the TMD synthesis and Janus conversion. In the case of MoSe$_2$, the synthesis-induced tensile strain is augmented by the additional tensile strain incurred during the Janus conversion which shatters Janus Mo$^S_{Se}$ crystals into small, 500-1000 nm across domains separated by physically few-nanometers-wide gaps. In contrast, when using MoS$_2$ as the precursor, the original tensile strain is compensated by the compressive strain upon Janus conversion which may lead to the formation of nano-wrinkles, with preserved continuity of Janus material. We demonstrated that the amount of the remaining compressive strain in Mo$^{Se}_S$, and consequently its topography, may be controlled by the choice of the original growth substrate. By substituting SiO$_2$/Si substrate with fused silica, we reduced the residual compressive strain in Janus Mo$^{Se}_S$ and successfully obtained almost wrinkle-free Janus Mo$^{Se}_S$ crystals.

We hope that the information obtained in the course of our study demonstrates the power of the cross-correlated AFM and TERS imaging of 2D materials and will be helpful for further optimization of the Janus conversion process. This would allow consistent synthesis of Janus monolayers free of cracks, bilayers, and wrinkles, as well as the control over the desired defects in these materials for specific applications.

## Methods

**AFM and TERS measurements:** AFM and TERS were conducted on a LabRam-Nano AFM-Raman system (HORIBA Scientific) equipped with 7 excitation lasers at 830, 785, 671, 633, 594, 532 and 473nm. Excitation and collection of TERS signal was done using the side 100×, 0.7 NA objective (Mitutoyo) inclined at 25º to the sample plane. Gold-coated or the Type II protected silver probes (HORIBA Scientific) based on Access-SNC AFM cantilevers (APPNano) were used for both the SPM and TERS characterization. Laser power past the objective for individual wavelengths was kept at the level of 100-250 µW. The colors in combined TERS maps that were representing the intensities of different Raman bands were adjusted individually for red, blue and green components without any strict protocol (i.e., red, blue, and green channels have their individual scale bars) in such a way that the structural features observed in corresponding combined maps were most clearly highlighted.

**Synthesis of Janus TMDs:** The starting pristine TMDs (MoS$_2$ and MoSe$_2$) were first grown on SiO$_2$ (300 nm)/Si and double-side polished fused silica substrates by CVD. The molybdenum precursor solution was prepared by mixing molybdenum trioxide (MoO$_3$) and potassium iodide (KI) in ammonia (NH$_4$OH), following a previously reported approach.[21] Subsequently, the solution was spin-coated onto substrates, and placed inside a one-inch tube furnace for sulfurization/selenization. The syntheses of MoS$_2$ and MoSe$_2$ were carried out at 750 ºC and 850 ºC for 5 min, respectively. For the MoS$_2$ synthesis, pure Ar (20 sccm) was applied as the carrier gas. In comparison, the MoSe$_2$ synthesis follows a hydrogen-free ramping (HFR)-CVD strategy that we recently developed.[22] In this process, pure Ar (50 sccm) was used during the temperature ramping stage, while 5 sccm of H$_2$ was only added during the synthesis stage when the temperature reached 850 ºC. The conversion of CVD-synthesized MoSe$_2$ and MoS$_2$ into Janus Mo$_{Se}^{S}$ and Mo$_{S}^{Se}$ follows the H$_2$ plasma-assisted ALS method.[6] The as-synthesized TMDs were placed in a tube reactor that is equipped with an inductively coupled plasma system at the upstream, and the distance between the sample and the edge of the plasma coil is optimized to be ~5-7 cm, depending on the type of starting material (MoSe$_2$ or MoS$_2$). Sulfur or selenium powder was located near the upstream edge of the plasma coil, providing chalcogen sources for the ALS reaction. During the conversion process, the H radicals in the H$_2$ plasma assist the conversion of TMDs to Janus structures. The ALS reaction is kept at room temperature, and the reaction time is 5-15 min.

**STEM characterization:** To prepare STEM specimens, MoSe$_2$ flakes were transferred onto a TEM grid (Quantifoil Cu grid). Atomic-resolution STEM imaging was performed with a probe-corrected Thermo Fisher Scientific Themis Z G3 60–300 kV S/TEM operated at 200 kV with a beam current of 20~30 pA and 25 mrad convergence angle. All the measurements were conducted at room temperature.

**Acknowledgements**

The synthesis and structural characterization of the Janus Mo$_{Se}^{S}$ and Mo$_{S}^{Se}$ was supported by the U.S. Department of Energy, Office of Science, Basic Energy Sciences, under award number DE-SC0020042. The optical characterization work is supported by the Air Force Office of Scientific Research under award number FA2386-24-1-4049. The work was performed in part through the use of MIT.nano's facilities. We thank Dr. Aubrey N. Penn (MIT.nano Characterization) for technical assistance. M.A.P. acknowledges the support of the Brazilian agencies CNPq and Fapemig.

Any opinions, finding, and conclusions or recommendations expressed in this material are those of the author(s) and do not necessarily reflect the views of the United States Air Force.

## Conflicts of interest

The authors declare the following competing financial interest(s): HORIBA Scientific is the manufacturer of the equipment used in this study. Collaboration with industry and academia is a part of A.K. job responsibilities. The authors declare no additional conflicts of interest.

# Supporting Information for

# Synthesis-related nanoscale defects in Mo-based Janus monolayers revealed by cross-correlated AFM and TERS imaging

Tianyi Zhang[1, #], Andrey Krayev[2, #], Tilo H. Yang[1], Nannan Mao[1], Lauren Hoang[3], Zhien Wang[1], Hongwei Liu[1], Yu-Ren Peng[1,4], Yunyue Zhu[1], Eleonora Isotta[5], Maria E. Kira[6], Ariete Righi[6], Marcos A. Pimenta[6], Yu-Lun Chueh[1,4], Eric Pop[3,7], Andrew J. Mannix[7], and Jing Kong[1]*

Tianyi Zhang, Tilo H. Yang, Nannan Mao, Zhien Wang, Hongwei Liu, Yu-Ren Peng, Yunyue Zhu, Jing Kong

Department of Electrical Engineering and Computer Science, Massachusetts Institute of Technology, Cambridge, Massachusetts 02139, United States

E-mail: jingkong@mit.edu

Andrey Krayev

HORIBA Scientific, Novato, California, 94949, United States

Lauren Hoang, Eric Pop

Department of Electrical Engineering, Stanford University, Stanford, California 94305, United States

Yu-Ren Peng, Yu-Lun Chueh

Department of Materials Science and Engineering, National Tsing Hua University, Hsinchu, 30013, Taiwan

Eleonora Isotta

Max Planck Institute for Sustainable Materials, Dusseldorf 40237, Germany

Maria E. Kira, Ariete Righi, Marcos A. Pimenta

Departamento de Física, Universidade Federal de Minas Gerais, Av. Antônio Carlos, 6627, Pampulha, Belo Horizonte, 31270-901, Minas Gerais, Brazil

Eric Pop, Andrew J. Mannix

Department of Materials Science & Engineering, Stanford University, Stanford, California 94305, United States

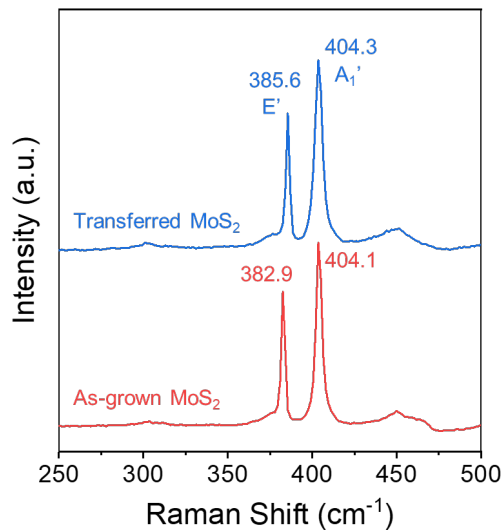

**Figure S1**. **Raman spectra of chemical vapor deposition (CVD)-grown monolayer $MoS_2$ on $SiO_2$/Si before and after the transfer process.** For this comparison, the as-grown $MoS_2$ was transferred to another piece of $SiO_2$/Si, which is identical to the substrate used for $MoS_2$ growth. After the transfer process which releases the growth-induced strain, the E' peak blue-shifted by ~2.7 cm$^{-1}$. According to the literature, the E' Raman mode shift rate as a function of biaxial strain is estimated as -4.48 cm$^{-1}$/%.[1] Thus, we can estimate that the tensile strain in as-grown $MoS_2$ is ~0.6%. Since the growth temperature and thermal expansion coefficient (TEC) of monolayer $MoSe_2$ is comparable to that of $MoS_2$, in **Figure 1**, we estimated the tensile strain in as-grown $MoSe_2$ as ~0.6% as well.

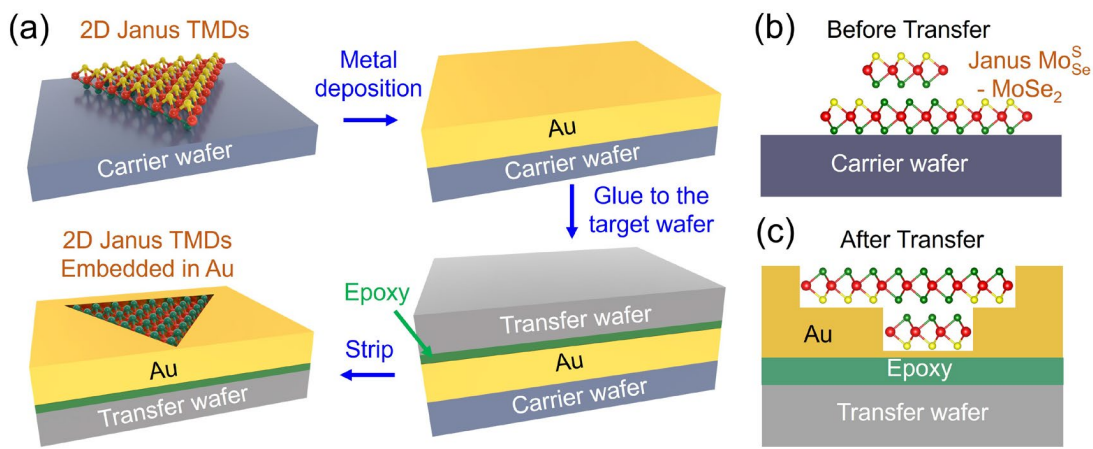

**Figure S2**. **Metal-assisted transfer approach for 2D Janus TMDs.** (a) A step-by-step illustration of the transfer of 2D Janus TMDs to noble metal substrates (e.g., Au, Ag) by using epoxy as a bonding layer. (b-c) Schematics of Janus $Mo^S_{Se}$-$MoSe_2$ vertical heterostructures before and after transfer. The transfer process leads to the heterostructures embedded in Au (or Ag) and vertically flips the crystal orientation.

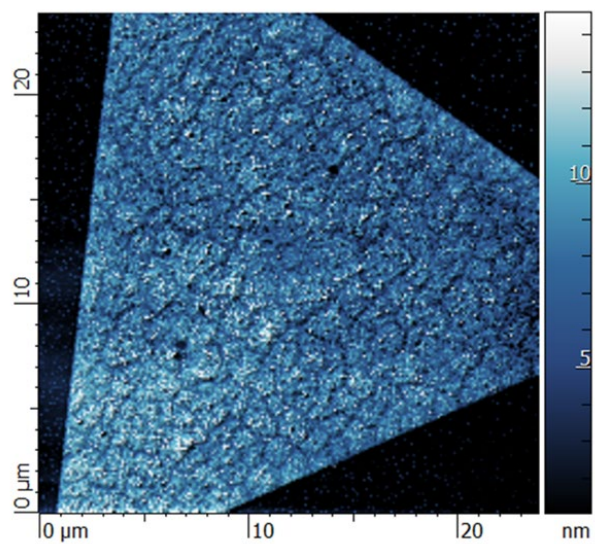

**Figure S3. Atomic force microscopy (AFM) topography image of as-synthesized Janus $Mo^S_{Se}$ crystal converted from $MoSe_2$.** As we can clearly see, the $Mo^S_{Se}$ crystal fragments into small domains.

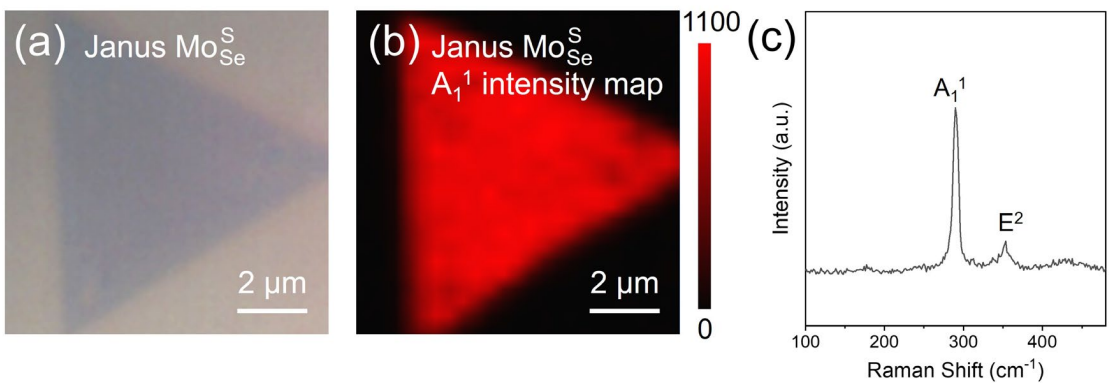

**Figure S4. Far-field optical characterization of Janus Mo$_{Se}^{S}$.** (a) Optical image, (b) the corresponding far-field Raman mapping ($A_1^1$ mode intensity), and (c) a typical Raman spectrum of Janus Mo$_{Se}^{S}$ extracted from the Raman mapping. In (b) and (c), the features of Janus Mo$_{Se}^{S}$-MoSe$_2$ vertical heterostructures are absent due to diffraction-limited spatial resolution.

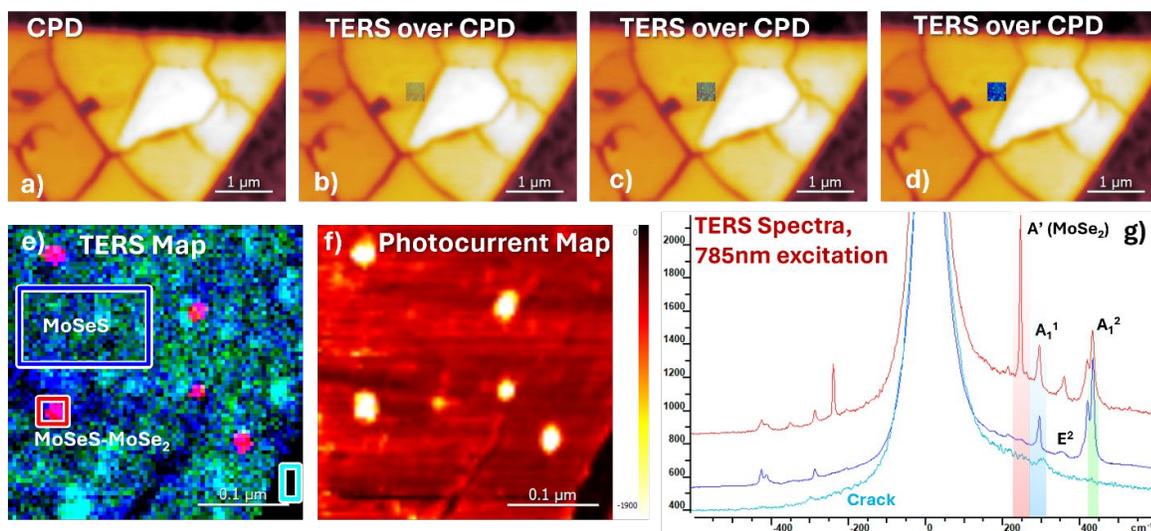

**Figure S5. Cross-correlated contact potential difference (CPD), tip-enhanced Raman spectroscopy (TERS), and photocurrent maps of Janus Mo$_{Se}^{S}$ converted from MoSe$_2$.** (a) CPD image of the transferred Janus Mo$_{Se}^{S}$ flake on the gold substrate. (b-d) TERS maps overlaid over the CPD image of Janus Mo$_{Se}^{S}$ with varied transparency. (e) The TERS map showing the intensity distribution of correspondingly highlighted Raman modes in panel (g), namely - MoSe$_2$ $A_1'$ (red), Mo$_{Se}^{S}$ $A_1^2$ (green), and Mo$_{Se}^{S}$ $A_1^1$ (blue). (f) Photocurrent map collected concurrently with the TERS map, showing a strong increase in the intensity of the photocurrent over the nanoscale islands of Mo$_{Se}^{S}$-MoSe$_2$ vertical heterostructures, as should be expected when the photocurrent value is limited by the light absorption. (g) The TERS spectra averaged over the correspondingly colored boxes in panel (e), which represent Janus monolayer Mo$_{Se}^{S}$ (blue), Mo$_{Se}^{S}$-MoSe$_2$ vertical heterostructures (red), and crack (cyan) regions.

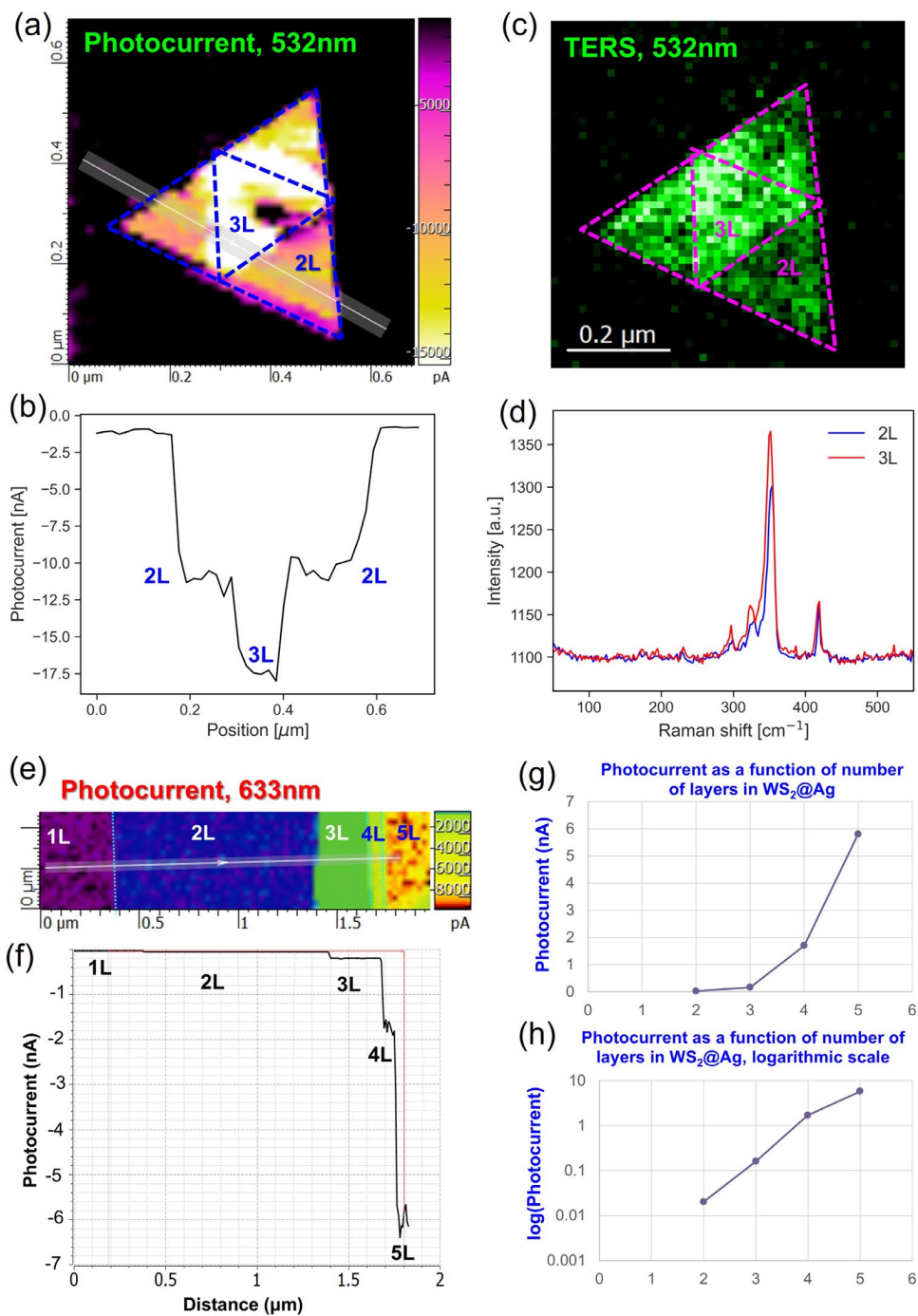

**Figure S6. Photocurrent data as a function of the layer number in WS$_2$@Ag for 532 nm and 633 nm excitation.** Photocurrent (a-b) and TERS (c-d) collected with excitation of 532 nm on a multilayer WS$_2$@Ag flake, and photocurrent data collected with 633 nm excitation (e-h) on another multilayer flake. As visible, the photocurrent increases with the layer number, which is attributed to the larger absorption possible with more layers, at least within the low layer number limit. The increase in photocurrent with layer number depends on the excitation energy too, and shows an almost exponential trend under illumination close to the A exciton of WS$_2$ (see e-h collected with 633 nm excitation wavelength).

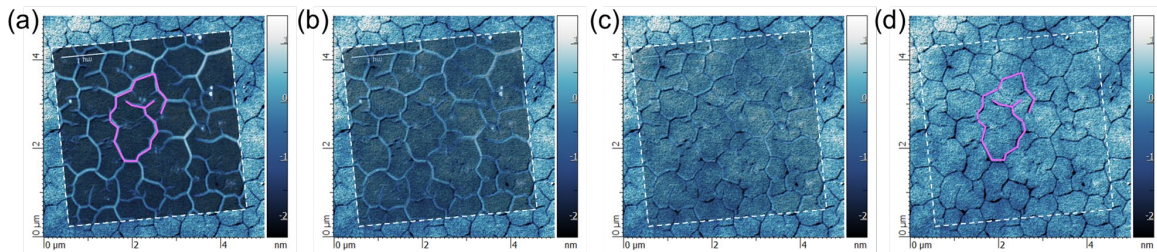

**Figure S7. Superimposed AFM topography images (with varied transparency) of as-grown Janus Mo$_S^{Se}$ on SiO$_2$/Si (flipped image) and transferred Janus Mo$_S^{Se}$ on silver.** We clearly see that the features on as-grown and transferred Janus Mo$_S^{Se}$ display a perfect match, confirming that the two AFM images were obtained in the same area. Additionally, the wrinkles in the as-grown crystal appears to be "cracks", but actually, they are inverted wrinkles after the silver-assisted transfer.

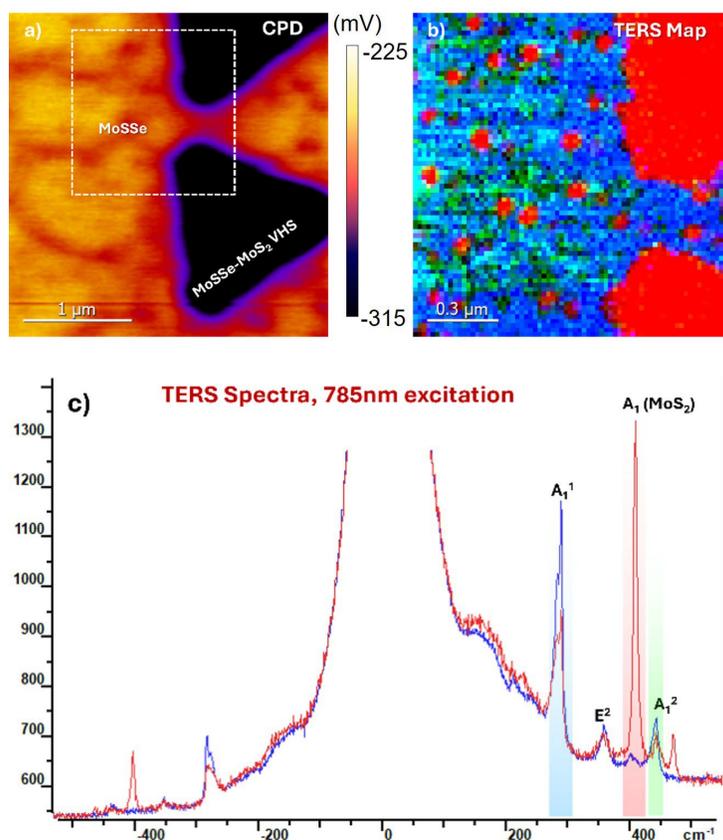

**Figure S8. CPD and TERS characterization of microscale and nanoscale islands of Janus Mo$_S^{Se}$-MoS$_2$ converted from MoS$_2$.** (a) CPD image of Janus Mo$_S^{Se}$ transferred onto silver, which captures over 1 μm sized islands of Janus Mo$_{Se}^{S}$-MoSe$_2$ vertical heterostructures. The area within the white dotted square was imaged using TERS. (b) The TERS map showing the intensity distribution of correspondingly highlighted Raman modes in panel (c), namely - MoS$_2$ A$_1$' (red), Mo$_S^{Se}$ A$_1^2$ (green), and Mo$_S^{Se}$ A$_1^1$ (blue). Nanoscale islands of Janus Mo$_{Se}^{S}$-MoSe$_2$, which are undetected by CPD in (a), are resolved by TERS mapping. (c) TERS spectra of Janus monolayer Mo$_S^{Se}$ (blue) and Mo$_S^{Se}$-MoS$_2$ vertical heterostructures (red).

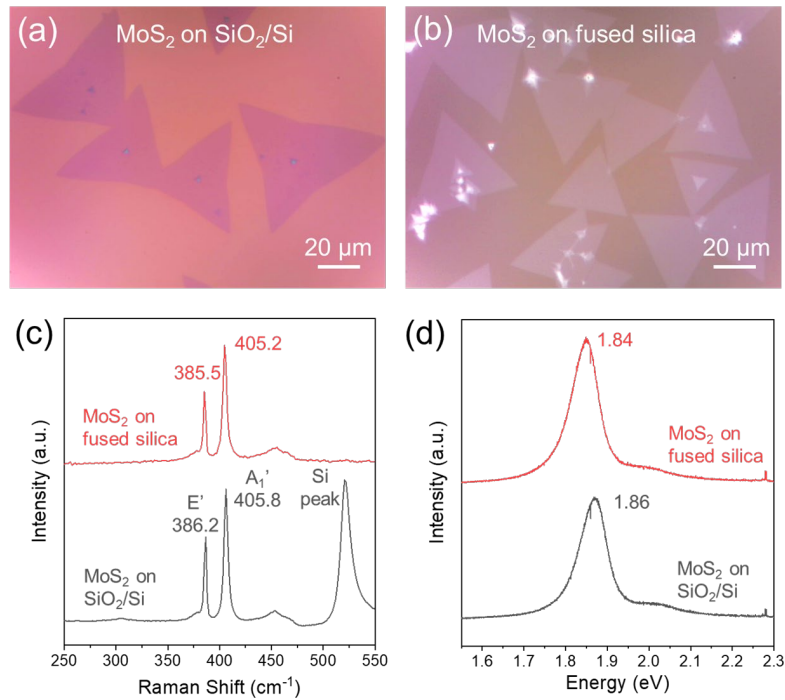

**Figure S9. A comparison between monolayer MoS₂ grown on SiO₂/Si and fused silica.** (a) Optical image of as-grown MoS₂ flakes on SiO₂/Si. (b) Optical image of as-grown MoS₂ flakes on fused silica. (c) Normalized Raman spectra and (d) Photoluminescence (PL) of MoS₂ grown on SiO₂/Si and fused silica. Red shifts in MoS₂ E' Raman mode and PL emission energy are observed in CVD-synthesized monolayer MoS₂ on fused silica, which indicates a higher amount of tensile strain than that of MoS₂ grown on SiO₂/Si.

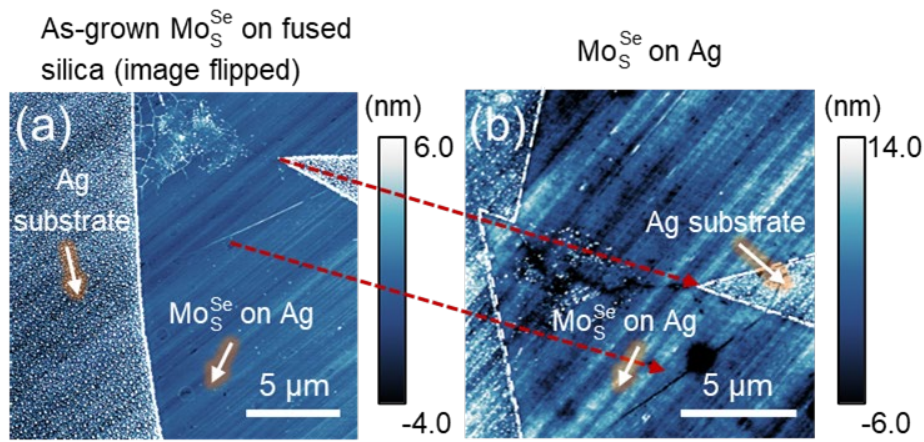

**Figure S10. AFM topography images of as-grown Janus Mo$_S^{Se}$ on fused silica (image is flipped to compare with the same region of Mo$_S^{Se}$ on Ag) and transferred Janus Mo$_S^{Se}$ on silver.** The morphological characteristics on the Janus Mo$_S^{Se}$ sample (marked by the red arrows) confirm that the same flake was measured before and after the transfer process. The edges of Mo$_S^{Se}$ in (b) are outlined by white dashed lines to differentiate Mo$_S^{Se}$ and bare silver substrate regions.

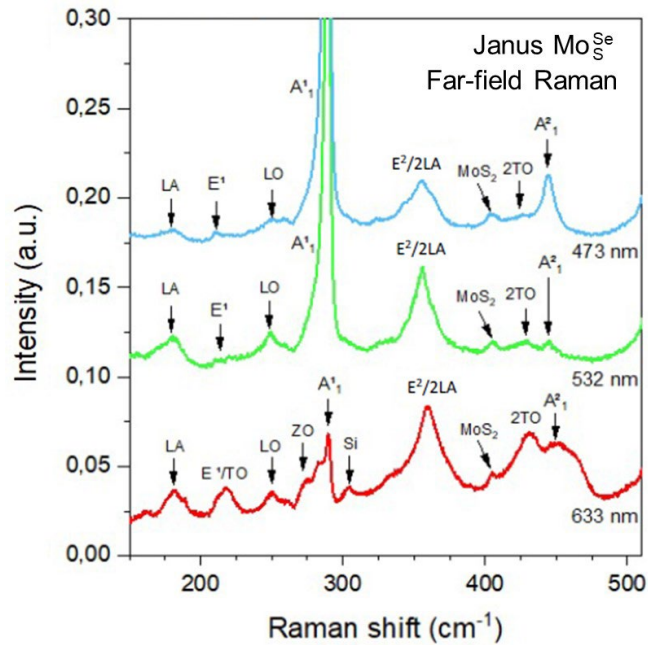

**Figure S11. Excitation wavelength-dependent far-field Raman spectra of as-grown Janus Mo$_S^{Se}$ on SiO$_2$/Si.** The relative intensity of different Raman modes (such as $A_1^1$ and $A_1^2$ modes) varies with excitation wavelengths.